\documentclass[twocolumn]{aastex63}

\received{June 1, xxxx}
\revised{January 10, xxxx}
\submitjournal{}
\shorttitle{}
\shortauthors{Wang et al.}

\usepackage{amsmath}
\begin{document}

\title{Avoiding the Geometric Boundary Effect in Shear Measurement}

\correspondingauthor{JUN ZHANG}
\email{betajzhang@sjtu.edu.cn}

\author{HAORAN WANG}
\affiliation{Department of Astronomy, Shanghai Jiao Tong University, Shanghai 200240, China}

\author{JUN ZHANG}
\affiliation{Department of Astronomy, Shanghai Jiao Tong University, Shanghai 200240, China}
\affiliation{Shanghai Key Laboratory for Particle Physics and Cosmology, Shanghai 200240, China}

\author{HEKUN LI}
\affiliation{Department of Astronomy, Shanghai Jiao Tong University, Shanghai 200240, China}

\author{ZHI SHEN}
\affiliation{Department of Astronomy, Shanghai Jiao Tong University, Shanghai 200240, China}

\begin{abstract}

In image processing, source detections are inevitably affected by the presence of the geometric boundaries in the images, including the physical boundaries of the CCD, and the boundaries of masked regions due to column defects, bright diffraction spikes, etc.. These boundary conditions make the source detection process not statistically isotropic. It can lead to additive shear bias near the boundaries. We build a phenomenological model to understand the bias, and propose a simple method to effectively eliminate the influence of geometric boundaries on shear measurement. We demonstrate the accuracy and efficiency of this method using both simulations and the z-band imaging data from the third data release of the DECam Legacy Survey. 
\end{abstract}


\section{Introduction} \label{sec:intro}

 For better understanding the physical origin of dark matter and dark energy, a number of ongoing and planned surveys are focusing on precisely measuring the weak lensing effect \citep{Abell2009,Laureijs2011,Hildebrandt2016,troxel18,Hikage2019}. Accurate measurement of the cosmic shear signals remains to be a challenge \citep{Hoekstra2008,Kilbinger2015,Mandelbaum2015a}. An important reason is that shear signals contain a wide range of systematic errors, leading to biased cosmological constraints. Currently, the study of systematic errors in shear measurement is still an important direction \citep{Hirata2003,Bernstein2010,Bridle2010,Voigt2010,Refregier2012,Kacprzak2014,Mandelbaum2014,Mandelbaum2015,Martinet2019,Sheldon2019,Pujol2020}. Some of the systematic errors are specific to the shear measurement method (e.g., model bias), and others are general problems for any weak lensing pipeline. In this work, we study the shear bias caused by the geometric boundaries in the images, a problem that we believe belongs to the second category. 

There are several different kinds of geometric boundaries existing in a typical CCD image, including the physical CCD boundaries, column defects due to the failure of charge transfer along the readout direction, and masked areas due to, e.g., the diffraction spikes of bright stars. For example, fig.1 shows a part of a typical CCD image with column defects from CFHTLenS \citep{Erben2009,Heymans2012,Erben2013}. As such boundary condition is locally anisotropic, it breaks the statistical isotropy of source detection/selection, leading to shear biases that are most significant near the boundaries. The purpose of this work is to give a phenomenological model for understanding this effect, and to propose a simple solution to avoid this type of bias. This is done in \S\ref{sec:style}. We then give a brief conclusion and discuss related issues in \S\ref{sec:highlight}.

\begin{figure}
	\centering
	\includegraphics[width=\columnwidth]{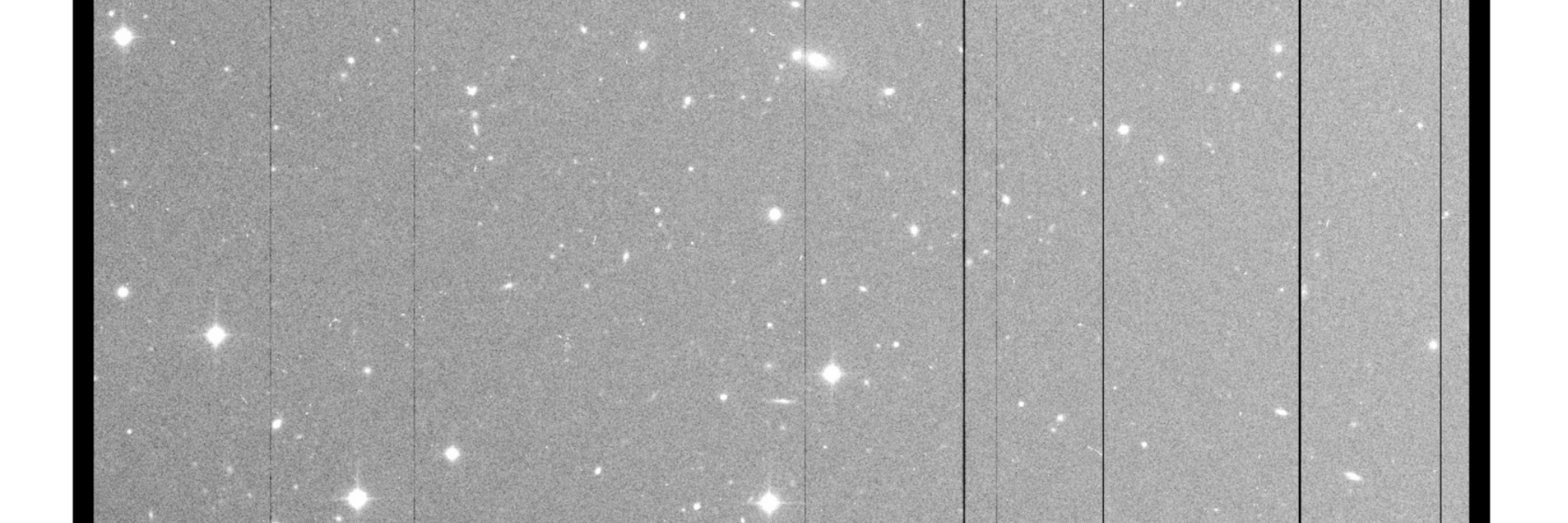}
	\caption{Bad columns in a typical CFHTLenS CCD image.}
\end{figure}

\section{Boundary problem and its solution} 
\label{sec:style}

Near the boundary, a galaxy is more likely to be detected as a valid source if its orientation is parallel to the boundary. This is demonstrated in fig.2, in which there are two groups of galaxies at two different distances from the boundary (marked by the dark thick line). The four galaxies of each distance group have the same size and shape, but different pointing directions. For the group that is closer to the boundary, it can be seen that the galaxy pointing perpendicular to the boundary (marked red) is not likely a valid source for shear measurement, as part of its shape information is missing\footnote{In the model fitting methods, either on single exposure or multiple ones, images with missing parts are often used. We argue in \S\ref{sec:highlight} that in this case, there are still related anisotropic effects.}. Only the other three sources of the same group survive. Statistically, this effect breaks the isotropy of the intrinsic galaxy shape, potentially leading to an additive bias in shear measurement. The more vertical bad columns an image has, the worse the problem is. Similarly, horizontal boundaries also lead to additive shear biases locally, but with an opposite sign. The effect is localized near the boundary, within a distance that is about the typical size of the galaxies. 
This type of error can in principle cause B-mode \citep{Schneider2002,Crittenden2002} in the shear field.

\begin{figure}
	\centering
	\includegraphics[width=\columnwidth,height=0.4\columnwidth]{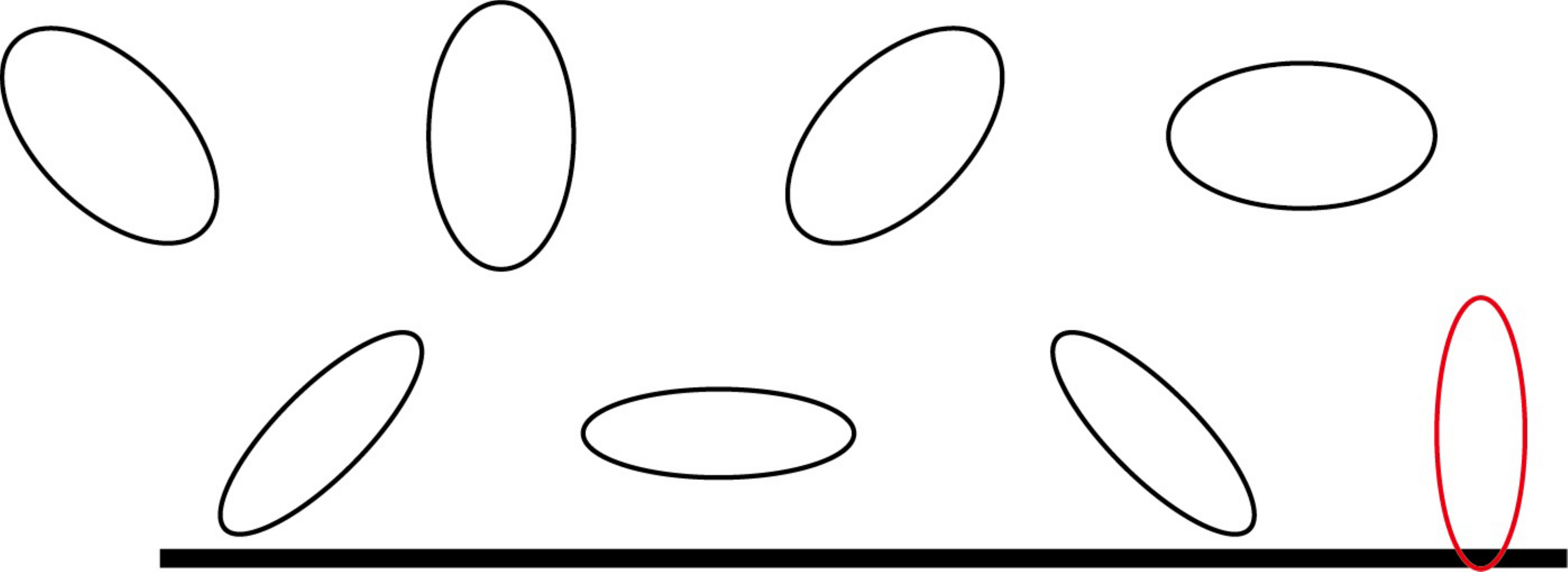}
	\caption{An example showing how the distribution of the galaxies near the boundary (the thick black line) become anisotropic. The galaxy of vertical orientation is more easily identified as a bad source (the red one). }
	\label{problem}
\end{figure}

\subsection{A simple model}
\label{model}
To understand the problem further, we build a model to analyze the boundary effect. Since the boundaries are mostly in the horizontal and vertical directions, this effect mainly affects the $g_{1}$ component. The area of the affected region is determined by the length of the boundary as well as the typical size of the galaxies. Let us assume that the total number of the observed galaxies on a CCD chip is $N$, among which $N^{+}$ and $N^{-}$ are sufficiently close to the horizontal and vertical boundaries respectively. On average, we expect a positive (or negative) additive shear bias $c$ (or $-c$) associated with every galaxy that is close enough to the horizontal (or vertical) boundaries. Thus we can get:

\begin{equation}
	g^{measure}_{1}=g^{true}_{1}+\frac{N^{+}-N^{-}}{N}\cdot c 
\end{equation}

For simplicity, let us consider the case of a rectangular CCD chip of size $l_x\times l_y$, without any bad columns in the middle. Let us also assume that the typical width of the affected regions along the boundaries is $s$, which should be comparable to the typical galaxy size. We also assume that the average galaxy number density is $n$. Then we have: $N^{+/-}\approx 2n\cdot s\cdot l_{x/y}$ and $N\approx n\cdot l_x\cdot l_y$, and therefore:

\begin{equation}
\label{gc}
	g^{measure}_{1}\approx g^{true}_{1}+2c\cdot \frac{s(l_x-l_y)}{l_xl_y}
\end{equation}
\subsection{Simulations}
\label{simulations} 

\begin{figure}
 	\centering
 	\includegraphics[width=\columnwidth,height=0.5\columnwidth]{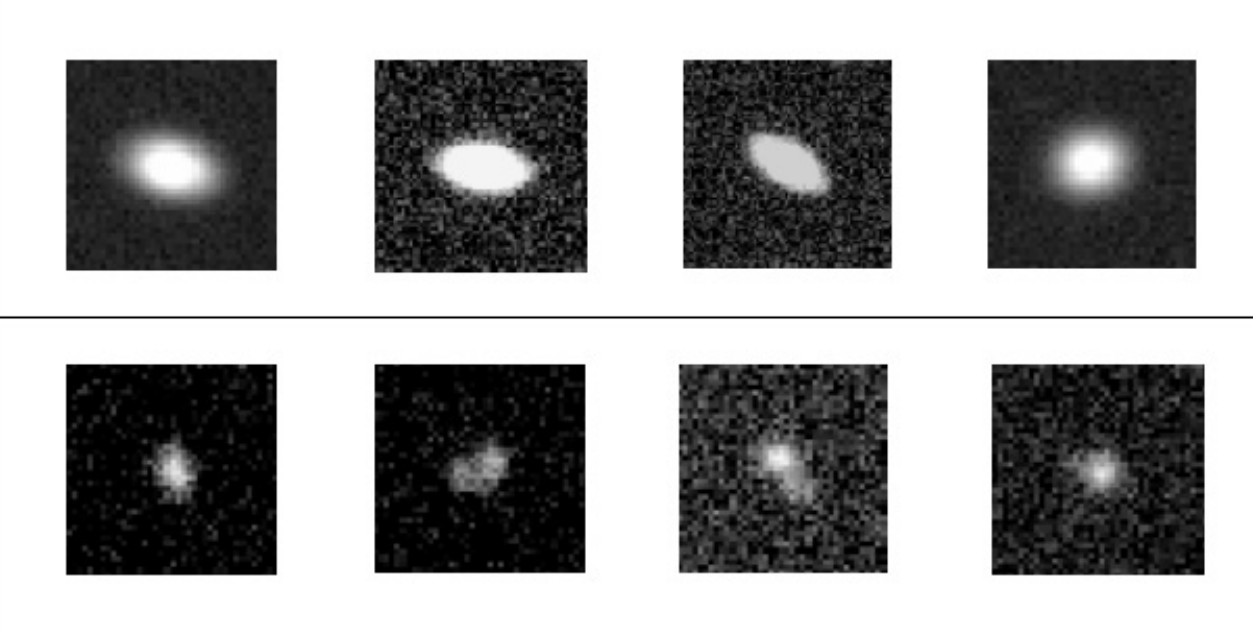}
 	\caption{The upper row: Galsim galaxies with regular shapes; the lower row: RW galaxies with irregular shapes.}
 	\label{comparasion}
\end{figure}

To demonstrate Eq.(\ref{gc}), we use two types of mock galaxies:  1) each galaxy is made of point sources whose positions are generated with the random walk algorithm proposed in \cite{Zhang2008} (called RW galaxy hereafter); 2) galaxies generated with the Galsim toolkit\citep{Rowe2015} (called Galsim galaxy hereafter). Galsim galaxies usually have more regular shapes than RW galaxies, as shown in fig.\ref{comparasion}.

In RW galaxy simulation, each galaxy consists of 60 point sources generated by 60 steps within a circular region of radius $r_s$ (9 pixels). The positions of these point sources are generated by the following algorithm:\\
\hspace{0.6cm}$\bullet$ The size of a single step is fixed at 1 pixel. The direction of each step is completely random. \\
\hspace{0.6cm}$\bullet$ The first step starts from the center of the circular region. Steps that are about to go beyond the region should restart from the center.\\
\hspace{0.6cm}$\bullet$ Each point source is assigned the same flux and convolved with the moffat-type PSF effect \citep{Bridle2009}. The pixel size of our simulations is set at 0.2 arcsec, and the FWHM of the PSF for the RW galaxies is 0.8 arcsec. \\

Our Galsim galaxies contain two kinds of profiles \citep{Simard2011}: the de Vaucouleurs profile and the exponential profile. For the buldge-dominated galaxies, which accounts for 10\% of the total galaxies, we adopt only the de Vaucouleurs profile. The rest disc-dominated galaxies are made as a combination of these two profiles. The probability distribution of the disc scale length is set as \citep{Miller2013}:
\begin{equation}
P(r)\propto r\exp{[-(r/a)^{\alpha}]},
\end{equation}
where $\alpha=4/3$, $a=r_{s}/0.833$, and $r_{s}$(arcsec) is decided by the $i$-band magnitude $m$ via $\ln(r_{s})=-1.145-0.269\times(m-23)$. The magnitude distribution is assumed to obey $P(m)\propto 10^{0.295m-1.08}$, and within the range of [20, 24]. We limit the radius $r$ of the galaxy in the range of [0.1, 0.4](arcsec). Each Galsim galaxy is also convolved with the moffat-type PSF, with the FWHM of 0.6 arcsec.

To study the boundary effect, we randomly place 2500 simulated galaxy imagies on a $5000\times5000$ grid, with a very low background noise added. The central $4000\times4000$ region is cut out to mimic a CCD chip. Each chip is divided into many sub-regions of a certain size (e.g., $200\times400$, $400\times800$), separated by masks, so that each sub-region can represent a geometrical boundary structure. For each boundary structure, we generate more than 30000 such CCD chips (each of which contains about 1600 randomly distributed galaxies). Five sets of shear values are applied to sources in these CCDs: (0.02, -0.02), (0.01, -0.01), (0.00,0.00), (-0.01, 0.01), (-0.02, 0.02). Each set of shear values therefore covers 6000 CCDs, containing around $10^{7}$ galaxies. We adopt the commonly used $m$ and $c$ to denote the shear bias as:
\begin{equation}
	g_{1,2}^{measured}=(1+m_{1,2})g_{1,2}^{input}+c_{1,2}
\end{equation}

\begin{figure}
	\centering
	\includegraphics[width=\columnwidth,height=0.85\columnwidth]{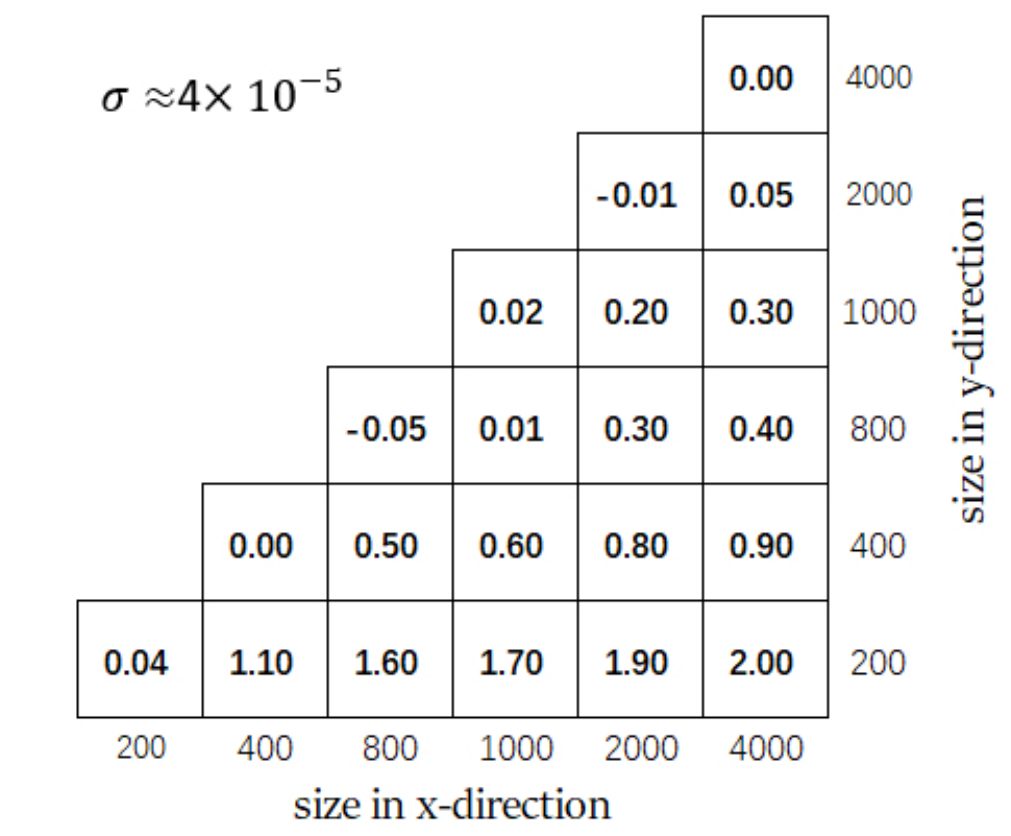}
	\caption{The additive bias $c_{1}$ for different boundary sizes ($l_x,l_y$). All $c_1$ values are in unit of $10^{-3}$. The variances are about $4\times 10^{-5}$. }
	\label{result1}
\end{figure}

We use the Fourier\_Quad pipeline for shear measurement \citep{Zhang2015,Zhang2016,Zhang2019}. The results for the RW galaxies are shown in fig.\ref{result1}, in which we list the values of additive bias $c_1$ as a function of the sub-region sizes along the x and y directions (i.e., $l_x$ and $l_y$)\footnote{$c_2$ is not shown, as it is consistent with zero in every case}. For symmetry reason, we only show the results for cases with $l_x\geq l_y$. We can see from the figure that for a fixed $l_y$, $c_{1}$ becomes larger when $l_x$ increases, and when $l_x$ is fixed, $c_{1}$ increases with a decreasing $l_y$. When $l_x=l_y$, $c_1$ is consistent with zero. These properties are all consistent with the prediction of our model in eq.(\ref{gc}). Note however that even in the case of $l_x=l_y$, there are still local additive shear biases near the boundaries. The overall null results are just due to cancellation globally.

\begin{figure}
	\centering
	\includegraphics[width=0.5\textwidth,height=1.4\columnwidth]{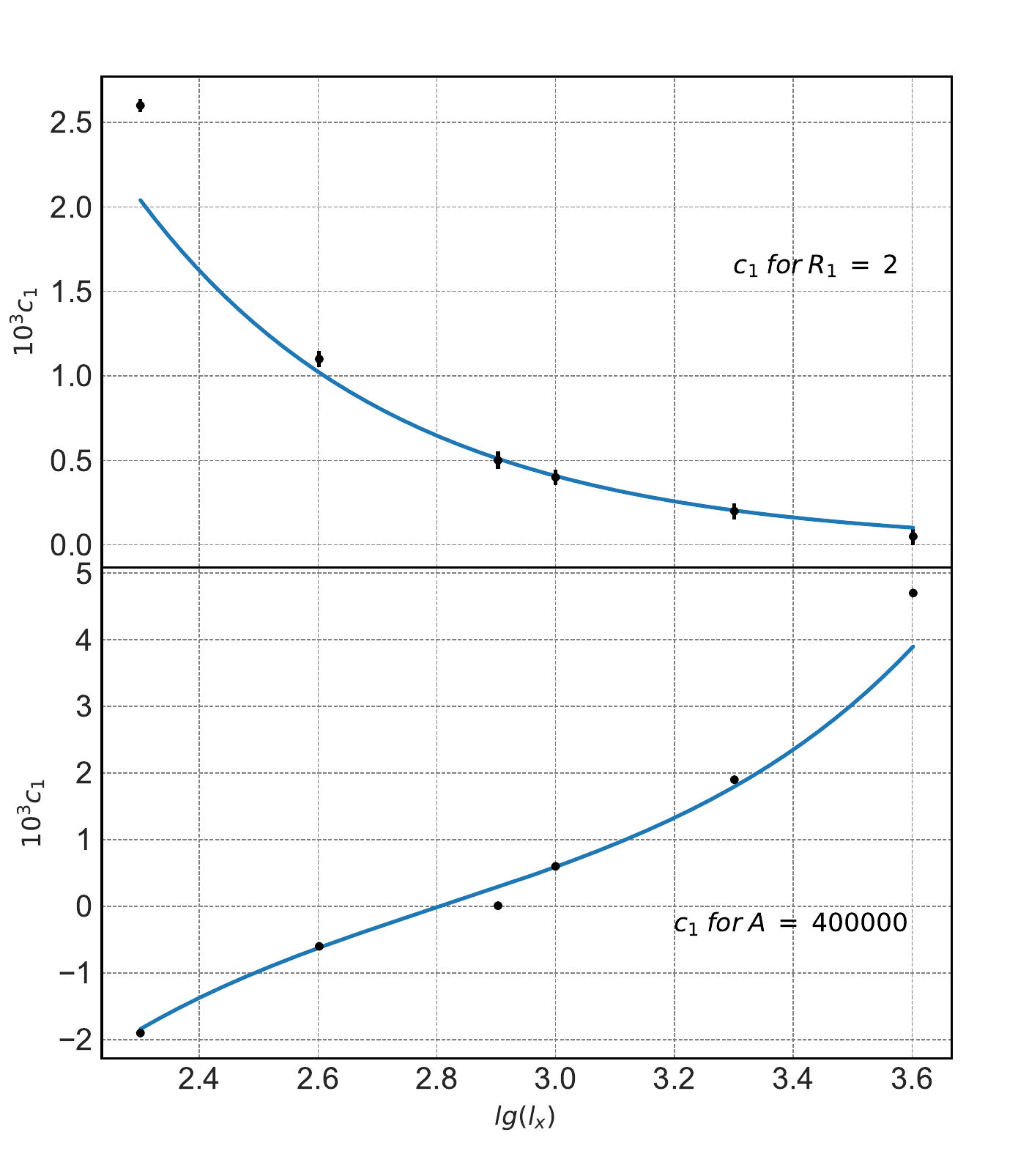}
	\caption{The change of the additive bias $c_1$ as a function of $l_x$. The upper pannel is under the condition of fixed axis-ratio ($l_x/l_y$), and the lower one is for fixed total-area ($l_x*l_y$). The solid curves are fittings from eq.(\ref{gc}) using the data shown in fig.\ref{result1}. The variances of the data points are about $4\times10^{-5}$.}
	\label{result1_2}
\end{figure}

The parameter $2cs$ defined in eq.(\ref{gc}) has a best-fit value of $408\pm 12$ in the current simulation. To further check the validity of the relation in the equation, one can quantitatively check the change of the bias $c_1$ with the ratio of the axis sizes $l_{x}/l_{y}$ for a fixed image area $l_{x}l_{y}$, or the other way around. The results are shown in fig.\ref{result1_2}, which are quite consistent with eq.(\ref{gc}). 

\begin{figure}
	\centering
	\includegraphics[width=0.56\columnwidth,height=0.48\columnwidth]{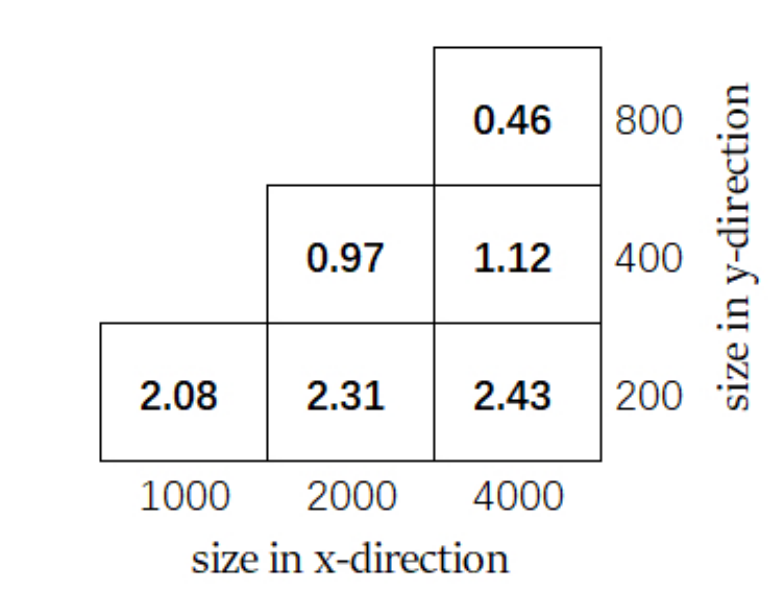}
	\caption{Similar to fig.\ref{result1}. The radii of the RW galaxies used here are larger than those used in fig.\ref{result1} by 50\%. All $c_1$ values are in unit of $10^{-3}$. The variances are about $4\times 10^{-5}$.}
	\label{result2}
\end{figure}

Another prediction of the model is that the boundary effect should be more significant for galaxies of larger average sizes. To observe this phenomenon, we generate another set of simulations with RW galaxies that are larger by 50\% than those in the previous set of simulations. The results are shown in fig.\ref{result2}, from which we can clearly see that $c_{1}$ is larger than those reported in fig.\ref{result1} under the same condition of ($l_x,l_y$).  

\begin{figure}
	\centering
	\includegraphics[width=0.50\columnwidth,height=0.45\columnwidth]{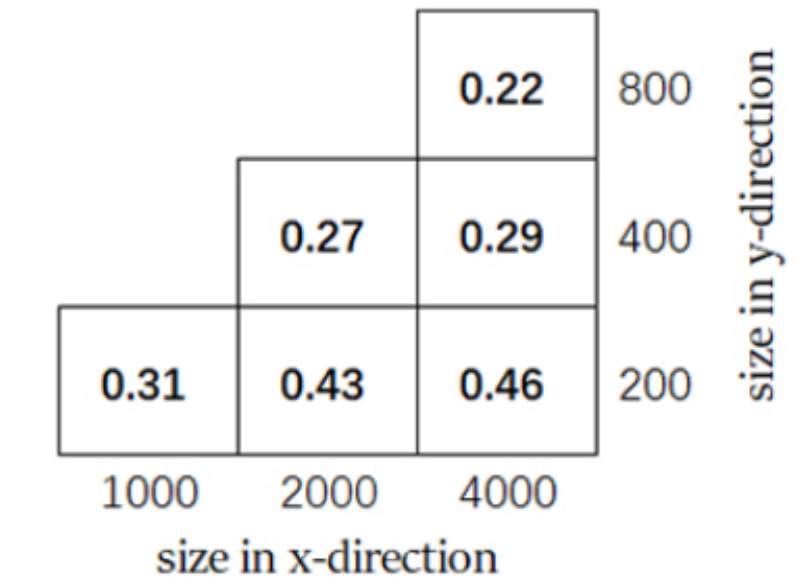}
	\caption{Similar to fig.\ref{result1}, but with Galsim galaxies. The ellipticity of these galaxies are set at $0.2$. All $c_1$ values are in unit of $10^{-3}$. The variances are about $1\times 10^{-5}$.}
	\label{e02}
\end{figure}

\begin{figure}
	\centering
	\includegraphics[width=0.50\columnwidth,height=0.45\columnwidth]{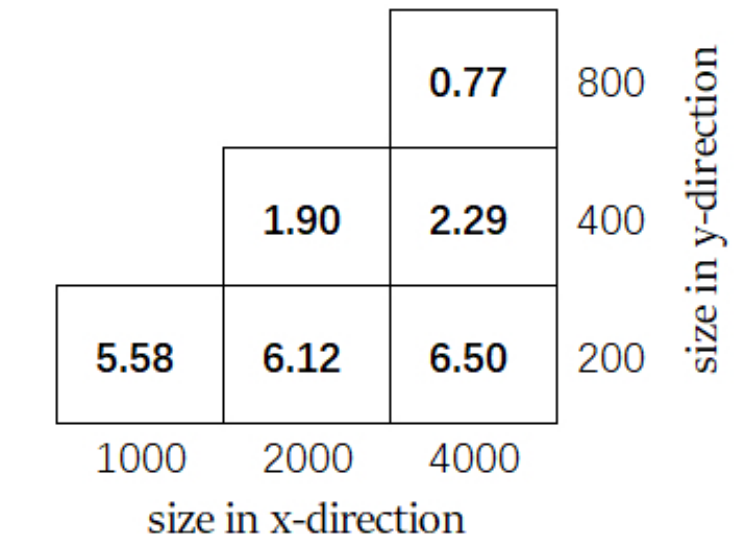}
	\caption{Same as fig.7, but with more elliptical Galsim galaxies. The ellipticity of the galaxies are set at $0.8$. All $c_1$ values are in unit of $10^{-3}$. The variances are about $1\times 10^{-5}$.}
	\label{e08}
\end{figure}

We also expect the boundary effect to be more significant for more elliptical galaxies. To observe this, we repeat the simulations twice with the Galsim galaxies, with the galaxy ellipticities set at $e=0.2$ and $e=0.8$ respectively. The results are shown in fig.\ref{e02} and fig.\ref{e08} accordingly. One can see that the $c_{1}$'s of the more elliptical cases are ten times larger than their counterparts with less elliptical sources.  

Overall, we can draw a conclusion that the additive bias due to the boundary is also influenced by the galaxy morphologies, such as their radii, ellipticities, etc.. The results in this section only represents the bias levels for several special cases.  

\subsection{The solution and its application on real data}\label{S}

\begin{figure}
	\centering
	\includegraphics[width=\columnwidth,height=0.4\columnwidth]{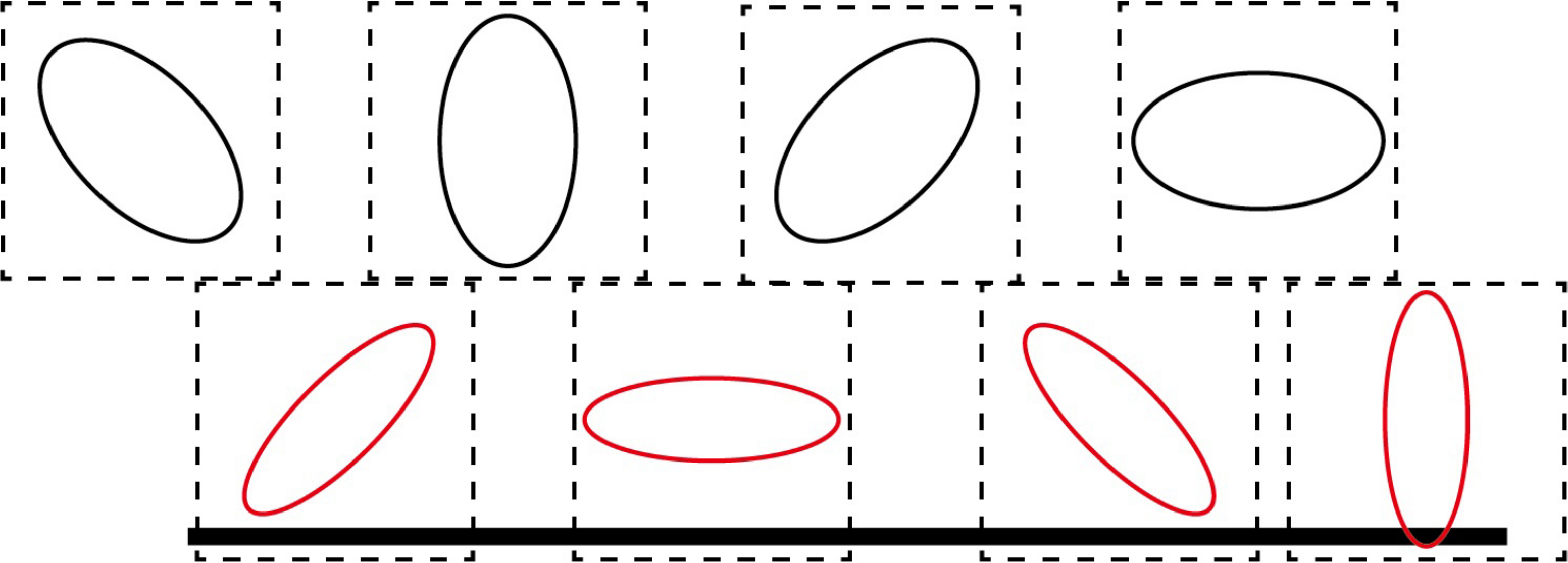}
	\caption{A way of avoiding the anisotropic source selection effect near the boundary is to remove the sources whose stamps are intercepted by the boundary. }
	\label{solu}
\end{figure}

As we have shown, source selection is a statistically anisotropic process near the image boundaries. The resulting shear bias cannot be treated as a universal additive bias, as it is significant only in the neighbourhood of the boundaries. To avoid such a bias, a natural solution is to require all selected sources to be at a certain distance away from the boundaries. It turns out that this condition can be easily realized by requiring that the boundaries do not intercept the image stamp. For example, in the case shown in fig.\ref{problem}, we would need to remove another three stamps that are close enough to the boundary, as shown in red in fig.\ref{solu}, although they are originally valid source images. 

\begin{figure}
	\centering
	\includegraphics[width=0.58\columnwidth,height=0.49\columnwidth]{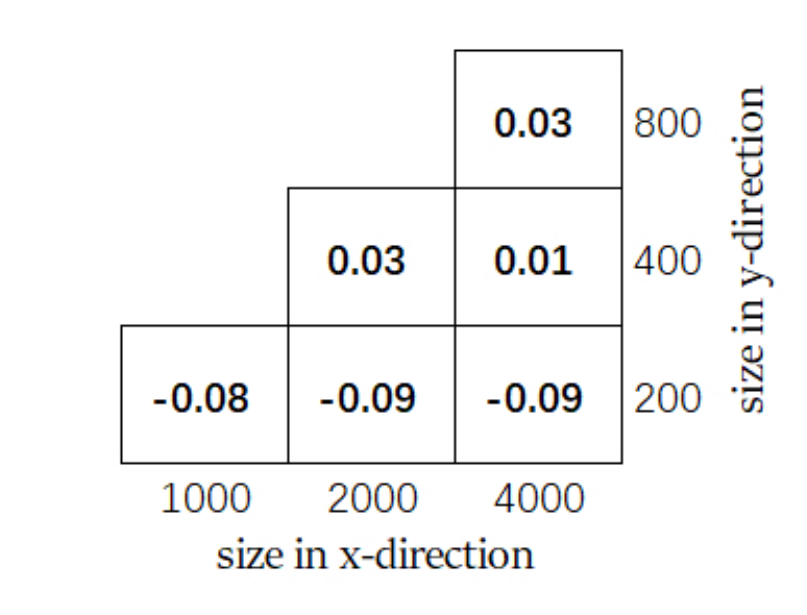}
	\caption{The new additive biases measured from galaxy sample in fig.\ref{result1} after the new source selection rule is applied. All $c_{1}$ values are in unit of $10^{-3}$. The variances are about $4\times 10^{-5}$ for all cases.}
	\label{result_solu}
\end{figure}

We find that this additional requirement is quite effective in removing the additive bias due to the existence of boundaries. For example, fig.\ref{result_solu} shows the new results of the simulations used in fig.\ref{result1} with the additional selection rule. The bias $c_1$ drops significantly from the level of $10^{-3}$ to $10^{-5}$. 

The validity of this solution can also be tested in real data. A direct way to do so is to adopt the idea recently proposed by \cite{Zhang2019}, which shows that one can use the galaxy shear estimators to restore the field distortion signal as a way of testing the shear recovery accuracy. It can be done within the observational data itself, without requiring simulations. As in \cite{Zhang2019}, here we again perform such a test with the Fourier\_Quad pipeline. Our imaging data is from the third data release of Dark Energy Camera Legacy Survey(DECaLS)\citep{Dey2019}. We use about 7400 calibrated z-band single exposures to do the test. Our source catalog is from \cite{Zou2019} (also see \cite{Zou2017}).

\begin{figure*}
	\centering
	\includegraphics[width=1.0\textwidth,height=\columnwidth]{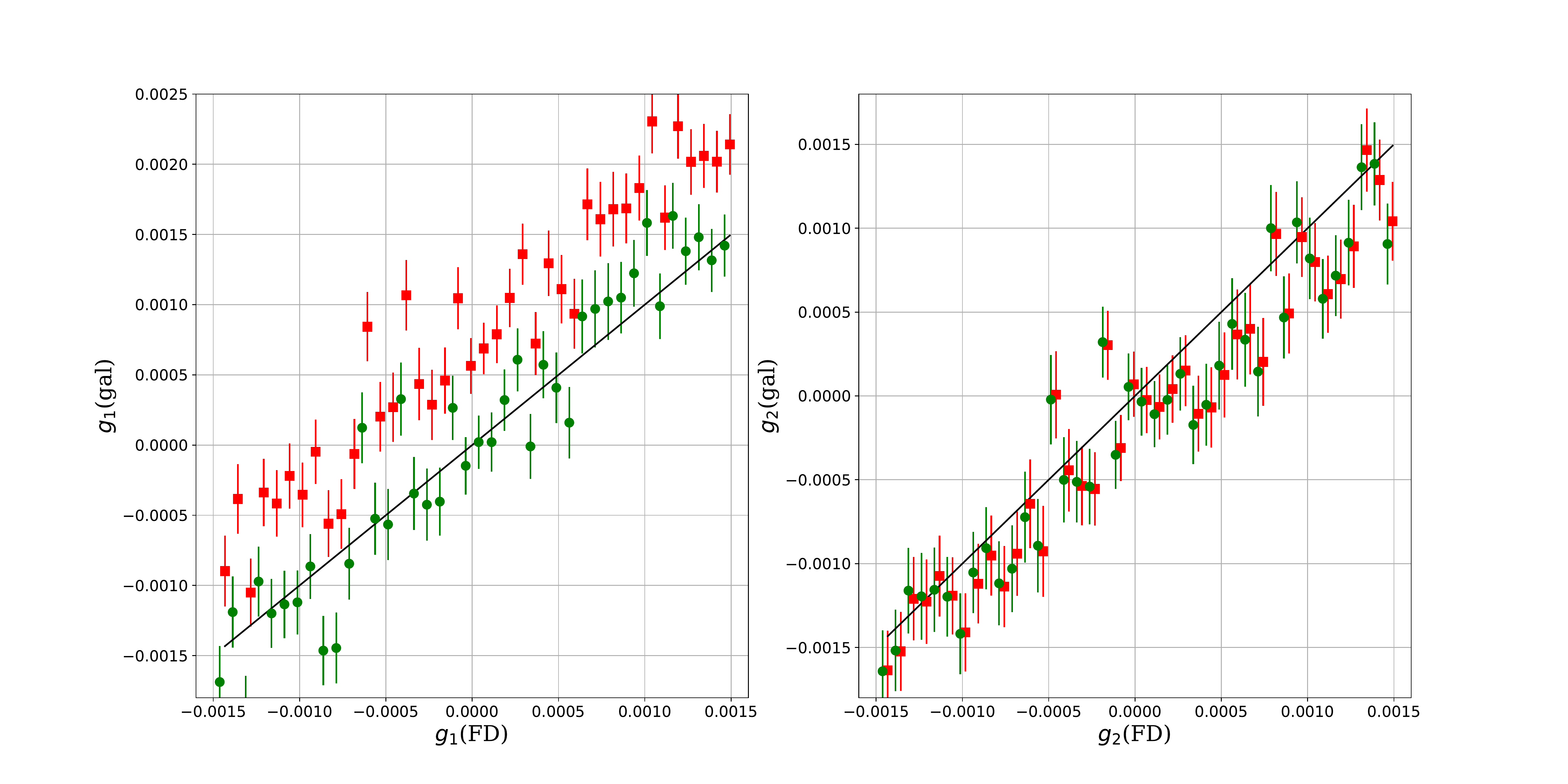}
	\caption{Test of shear recovery with the field-distortion signals using the z-band data of the third DECaLS Data Release. The left and right panels are the results of $g_1$ and $g_2$ respectively. The green and red points with $1\sigma$ error bars are the results with and without removing boundary effect respectively.}
	\label{DR}
\end{figure*}

In fig.\ref{DR}, we show the comparison between the galaxy shear and the signal induced by field distortion. The solid black line is "y=x", shown as a reference. The red square data points are the results from all valid sources confirmed by the Fourier\_Quad pipeline on single exposures. In this case, one can clearly see a positive additive bias. This is due to the fact that there are many more boundary effects acting along the x-axis (defined to be the direction of RA, which almost coincides with the direction of the longer sides of the CCD chip) than on the y-axis. The green circular points are the results achieved after adopting the new selection criterion regarding the boundary effect. This procedure remove about 5 - 10\% of the sources. We can see that by removing sources that are close enough to the boundaries, the additive bias in $g_1$ can indeed be suppressed, and the results of $g_2$ are hardly affected.   

\section{Conclusion and Discussions} \label{sec:highlight}

In this paper, we have specifically studied the geometric boundary effect on the accuracy of shear measurement. The existence of boundaries due to either the CCD edges or bad columns within the images can cause locally anisotropic source detection, thereby biasing the shear measurement. This is simply because sources are more likely to be detected/selected when its orientation is parallel to the boundary. It is a selection effect related to source detection. It causes local additive shear biases that are hard to remove globally. 

We propose a model to parameterize the shear bias as a function of the side lengths of the bounded region, and find a good agreement between the model predictions and the simulation results. In our simulations, we use a large number of mock galaxies of different morphological types, including both RW galaxies and Galsim galaxies. The results of the test as well as the model prediction agree with our intuition: the boundary effects are more serious when: 1. the area of the bounded region is small; 2. The ratio of the side lengths is large; 3. the galaxies have large sizes on average; 4. the galaxies are more elliptical. In all the above cases, the boundaries have more impact on the selection of the galaxy sample, and on the resulting shear bias. 

We find that a simple way of removing such a bias is to remove the sources that are too close to the boundaries even though their isophotes of the detection threshold (e.g., 2 - 4$\sigma$ above the background) do not yet touch the boundaries. This is done by requiring that no boundary marks (e.g., CCD edges, bad columns, etc.) cross the source stamp (of a predetermined fixed size). We use simulations to show that this simple procedure can indeed remove the additive shear bias related to the boundaries. We further demonstrate the existence of the boundary effect in real data, the DECaLS z-band imaging data, using the field-distortion test method proposed in \cite{Zhang2019}. The results also show that our treatment can accurately remove the bias from the boundary effect.  

We note that the isotropy of the source selection algorithm is also affected by the shear signal itself: two neighbouring sources are prone to overlap with each other when shear is along the line connecting them \citep{Sheldon2019}. This type of effect leads to multiplicative shear biases, in contrast to the additive biases addressed in this paper. We plan to study this effect in a future work.

Since the Fourier\_Quad pipeline processes shear measurement on individual exposures independently, our discussion regarding the boundary effect in this paper is quite straightforward and specific. In other shear measurement methods, e.g., model-fitting algorithms, it is often the case that multiple exposures are used together to construct the shear estimator for each galaxy. Images with missing pixels (due to boundaries for example) are often tolerated, and included in the fitting. If the images intercepted by the boundary marks are included in the fitting, their anisotropic missing-pixel distribution can still bias the shear estimators in principle. On the other hand, if such contaminated images are discarded in the fitting, the survived ones on the same exposure would tend to lie along the boundaries when they are near the boundaries, thereby causing shear bias as well. In the second case, one can consider adopting the treatment in this paper: discarding the image according to whether the boundary marks cross its stamp, i.e., a domain of a fixed size, instead of the isophote of the detection threshold. A detailed study of this topic is however beyond the scope of this work.

\newpage
\acknowledgments

We thank Xiaokai Chen and Dezi Liu for their help in galaxy simulation and shear measurement. The computations in this paper were run on the $\pi$ 2.0 cluster supported by the Center for High Performance Computing at Shanghai Jiao Tong University.

This work is supported by the National Key Basic Research and Development Program of China (No.2018YFA0404504), and the NSFC grants (11673016, 11621303, 11890691, 12073017).

\newpage
\bibliography{0602}{}
\bibliographystyle{aasjournal}

\end{document}